\documentclass[aps,prb,preprint,showpacs,showkeys,floatfix]{revtex4}

\usepackage{graphicx,color}
\graphicspath{{figs/}}
\begin{document}
\title{Mechanical Oscillation of Kinked Silicon Nanowires: a Natural Nanoscale Spring}
\author{Jin-Wu Jiang}
    \affiliation{Institute of Structural Mechanics, Bauhaus-University Weimar, Marienstr. 15, D-99423 Weimar, Germany}
\author{Timon Rabczuk}
    \affiliation{Institute of Structural Mechanics, Bauhaus-University Weimar, Marienstr. 15, D-99423 Weimar, Germany}

\date{\today}
\begin{abstract}
We perform classical molecular dynamics simulations to demonstrate the application of kinked silicon nanowires (KSiNWs) as nanoscale springs. The spring-like oscillation in gigahertz frequency range is successfully actuated using a similar procedure as the actuation of a classical mass spring oscillator. We detect the spring-like mechanical oscillation and some other low-frequency oscillations by the energy spectrum analysis, where a dimensional crossover phenomenon is observed for the transverse mode in KSiNWs with decreasing aspect ratio. Our findings shed light on the elastic properties of the KSiNW and open a way for its application in nanomechanical devices.
\end{abstract}

\maketitle

The kinked silicon nanowire (KSiNW) was synthesized by Tian {\it et al.} in Lieber's group at Harvard University in 2009~\cite{TianB2009}. Owing to its peculiar kinking structure, this nano structure is viewed as one of the promising building blocks in bottom-up integration of active devices. It is able to manipulate the arm length of the kink by controlling the growth time. The change of growth direction at kinks is the characteristic of KSiNWs; thus great experimental efforts have been devoted to investigating the growth mechanism of the kinks in silicon nanowires\cite{ChenH,YanC,KimJ,PevznerA,ShinN,MusinIR}. Besides these experimental works, Schwarz and Tersoff have proposed a theoretical model to interpret the growth mechanism of the KSiNW~\cite{SchwarzKW,SchwarzKWprl}. In their model, the kink formation is attributed to the interplay of three basic processes: facet growth, droplet statics, and the introduction of additional facets. Stimulated by KSiNWs, several groups have synthesized kinks in other nanowires, such as ${\rm In_{2}O_{3}}$ multikinked nanowires~\cite{ShenG}, kinked germanium nanowires~\cite{KimJH}, germanium–silicon axial heterostructure with kinks~\cite{DayehSA}, and kinked ZnO nanowires~\cite{LiS}. As long as significant achievement has been gained for the kink growth mechanics, a meaningful task would be the investigation of possible practical applications of such material. A recent experiment has demonstrated the application of KSiNWs as nanoelectronic bioprobes.\cite{XuL} Hereby, we demonstrate its utility as a nanoscale spring in mechanical device.

In this paper, we study the mechanical oscillation of KSiNWs of various aspect ratio by molecular dynamics (MD) simulation. We demonstrate a successful actuation of the spring-like oscillation in the KSiNW by mimicking the actuation procedure for a classical mass spring system. We perform an energy spectrum analysis for the mechanical oscillations, where the spring-like oscillation and some other low-frequency oscillation modes are successfully assigned to corresponding resonant peaks in the spectrum. From energy spectrum analysis, we also find the dimensional crossover behavior for the transverse oscillation mode in KSiNWs with decreasing aspect ratio.

\begin{figure}[htpb]
  \begin{center}
    \scalebox{1.0}[1.0]{\includegraphics[width=8cm]{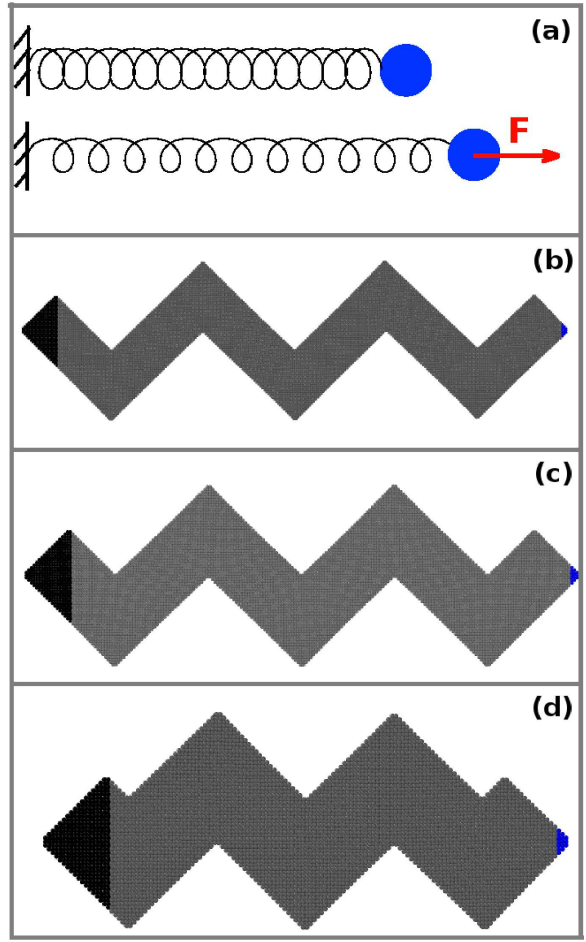}}
  \end{center}
  \caption{Configuration of the three KSiNW samples in present work. (a) The equilibrium position of a spring connected to a point mass. The right end is stretched by an external force $\vec{F}$ (with fixing left end). This mass spring oscillator starts to oscillate after $\vec{F}$ is removed. From (b) to (d): the three studied KSiNW samples S1, S2 and S3. Left end is fixed. Tips on the right ends are in blue online. Cartesian coordinate system: $x$ is along axial direction from left to right, and $z$ is the viewing direction of structures shown here.}
  \label{fig_cfg}
\end{figure}

Fig.~\ref{fig_cfg}~(b), (c), and (d) show the configuration of three samples in present study: sample 1 (S1), sample 2 (S2), and sample 3 (S3). In all samples, the nanowire growth direction changes from one $<100>$ direction to another $<100>$ lattice direction, resulting in a kinking angle of 90$^{\circ}$. All of the four side surfaces are the $\{100\}$ lattice surface, enclosing a square cross section. The cross section is of $27\times 27$~{nm$^{2}$}. For S1, S2, and S3, the aspect ratio (length to thickness ratio) is 3.6, 3.0, and 2.4; while the number of silicon atoms is $8\times 10^{4}$, $6\times 10^{4}$, and $4\times 10^{4}$.

All MD simulations were performed using the publicly available simulation code LAMMPS~\cite{PlimptonSJ}, while the OVITO package was used for visualization~\cite{ovito}. The interaction between silicon atoms is described by the Stillinger-Weber potential~\cite{StillingerF}. Owning to its efficiency and accuracy, this empirical potential has gained a wide application in the simulation of silicon and other similar valence bonded systems. The Newton equations of motion are integrated using the velocity Verlet algorithm with a time step of 1 fs. In the growth direction, the left end (black online) is fixed in all simulations, while the right end is free. Free boundary condition is applied in lateral directions.

To actuate the mechanical oscillation of KSiNWs, we first learn a lesson from the classical approach to generate the oscillation of a harmonic mass spring oscillator shown in Fig.~\ref{fig_cfg}~(a). The point mass on the right end is displaced away from its equilibrium position $l_{0}$ by an external force $\vec{F}$ to an updated position $l$. The system starts to oscillate harmonically after releasing $\vec{F}$. The oscillation amplitude is $A=l-l_{0}$. The maximum strain inside the spring during the oscillation is $\epsilon_{m}=A/l_{0}$.

In a similar way, we actuate the mechanical oscillation of the KSiNW in following four steps. Firstly, the structure is relaxed to the minimum energy state using the conjugate gradient algorithm. Secondly, the system is stretched gradually by adding a ramp displacement that scaled linearly from zero at the fixed left end to a maximum value at the free right end. The displacement incremental results in a strain rate of $\dot{\epsilon}=10^{9}$~s$^{-1}$, which is a typical value in MD simulations~\cite{JiangJW2012jmps}. This displacement incremental procedure is stopped after the tension inside the system reaches the required value, $\epsilon_{m}$. In following simulations, $\epsilon_{m}=$ 0.01 or 0.02, which are small enough to avoid plastic deformation in the KSiNW. Thirdly, the right tip (blue online) in the tensile system is clamped, and the system is thermalized by the Nos\'e-Hoover\cite{Nose,Hoover} thermostat to a constant temperature of 4.2~{K} within the NVT (i.e. the particles number N, the volume V, and the temperature T of the system are constant) ensemble for 20~{ps}. We focus on the liquid helium temperature because this temperature is commonly utilized in experiments involving nanoscale mechanical oscillations (eg. 90~{mK} or 4.0~{K} in Ref.~\onlinecite{EichlerA}). Fourthly, the right tip is set free and the system is allowed to oscillate freely within the NVE (i.e. the particles number N, the volume V, and the energy E of the system are constant) ensemble for 4~{ns}. The trajectory and energy time history from the final step are then used in the analysis of the mechanical oscillation for the KSiNW.

\begin{figure}[htpb]
  \begin{center}
    \scalebox{1.0}[1.0]{\includegraphics[width=8cm]{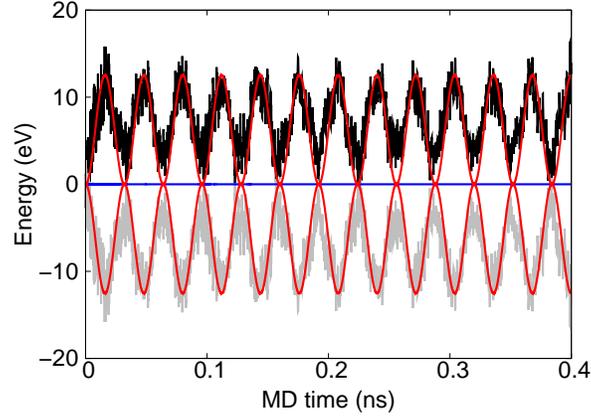}}
  \end{center}
  \caption{Energy oscillation due to the mechanical oscillation in KSiNW S1. Energies are with reference to the value just before the actuation of the mechanical oscillation (i.e t=0). The curve above (bellow) zero is the kinetic (potential) energy. The energy oscillates between kinetic and potential energy at frequency $f=31.26$~{GHz}, corresponding to the mechanical oscillation. The total energy is the horizontal line at zero, demonstrating a good energy conservation in the NVE ensemble. Two cosine functions (red online) are used as guides to the eye. We note that the energy oscillation frequency is double of the mechanical oscillation frequency.}
  \label{fig_energy}
\end{figure}

Fig.~\ref{fig_energy} shows the energy time history in KSiNW S1 during the simulation within NVE ensemble, with $\epsilon_{m}=0.01$. All energies are with reference to their value at $t=0$, i.e at the very beginning of the step 4 in the above actuation procedure. The kinetic (potential) energy is above (bellow) the $x$ axis; while the total energy locates exactly at the $x$ axis. Corresponding to the mechanical oscillation, the energy oscillates between kinetic and potential energy, and an excellent energy conservation has been achieved. This nice energy oscillation behavior exhibits a good spring-like behavior of the KSiNW. We note that the frequency of the energy oscillation ($f=31.26$~{GHz}) is twice of the mechanical oscillation frequency. Two cosine functions (red online) are also shown as guides to the eye.

\begin{figure}[htpb]
  \begin{center}
    \scalebox{1.0}[1.0]{\includegraphics[width=8cm]{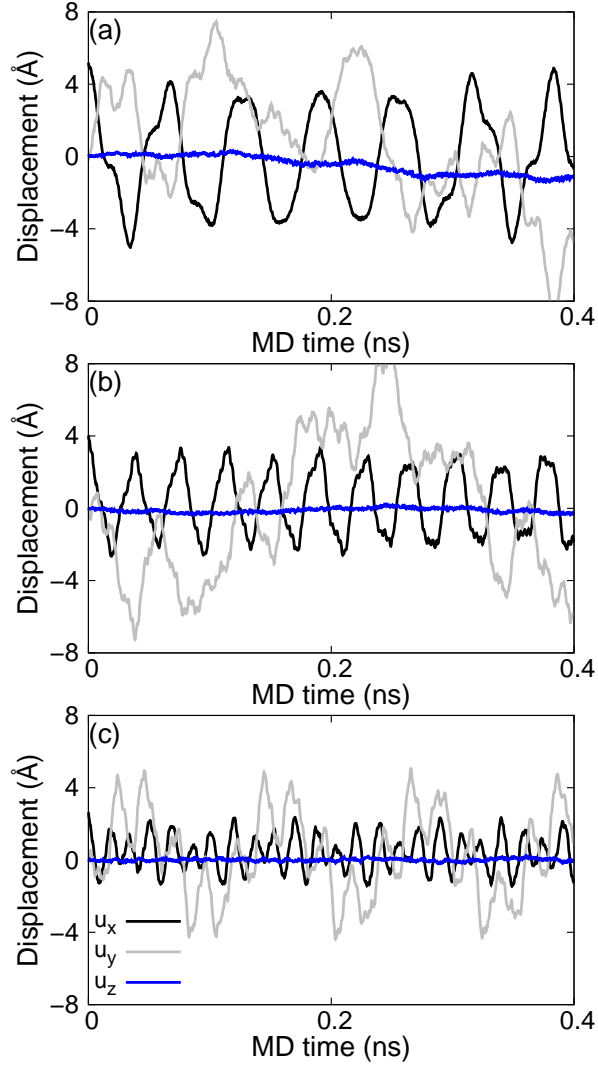}}
  \end{center}
  \caption{Vibrational displacements of the tip at the right end in KSiNWs. Top to bottom: results for S1, S2, and S3. $u_{x}$, $u_{y}$, and $u_{z}$ are the vibration in three Cartesian directions.}
  \label{fig_u}
\end{figure}

For quantitative analysis of the spring-like oscillation, we record the vibrational displacement of the right tip $\vec{u}(t)=(u_{x},u_{y},u_{z})=\vec{r}(t)_{\rm tip}-\vec{r}_{\rm tip}^{0}$. $\vec{u}$ is averaged over all silicon atoms within the tip on the right end. These vibrational displacements are plotted in Fig.~\ref{fig_u}~(a), (b), and (c) for the three studied samples S1, S2, and S3. The actuation strain here is $\epsilon_{m}=0.01$. In all samples, a steady oscillation in $x$ direction has been successfully actuated. This oscillation reflects the spring-like behavior of the KSiNW. The oscillation in $z$ direction only induces some neglectable fluctuations. However, we can observe some obvious oscillations in the $y$ direction, although our original goal is to actuate oscillation solely in $u_{x}$ by displacing atoms in the $x$ direction. This is due to the so-called `artificial' effect as discussed in Ref.~\onlinecite{JiangJW2012jap}. In that work, it was found that lots of other oscillation modes will be excited simultaneously if the mechanical oscillation is not actuated properly according to its fundamental vibration modes. In present work, the mechanical oscillation is actuated by a linearly ramp displacement. Due to surface and finite size effects, the linear distribution shape does not correspond to the fundamental vibration mode in the KSiNW, so the artificial effect leads to the excitation of lots of other oscillations besides the spring-like mechanical oscillation. From the $u_{x}$ curve in Fig.~\ref{fig_u}~(a), the mechanical oscillation frequency is extracted to be $f=15.63$~{GHz}, which is exactly half of the frequency in the energy oscillation in Fig.~\ref{fig_energy}. It indicates that the spring-like oscillation in $x$ direction makes most important contribution to the oscillation in the kinetic/potential energy, even though its oscillation amplitude is somehow smaller than the oscillation in $u_{y}$. We point out some roughness on both $u_{x}$ and $u_{y}$ curves, which are actually due to higher-frequency oscillation modes.

From Fig.~\ref{fig_u}, we have analyzed the spring-like oscillation based on the vibrational displacement. In this analysis, we have encountered some disruption induced by lots of other oscillation modes resulting from the artificial effect. The artificial effect can be eliminated by actuating the mechanical oscillation complying with the fundamental vibration mode. However, it is impossible to get the fundamental vibration mode for those KSiNWs here, because their lattice dynamics matrix can't be solved due to huge degrees of freedom. On the other hand, the oscillation in kinetic/potential energy shown in Fig.~\ref{fig_energy} disclose nicely the spring-like mechanical oscillation. Further more, a direct analysis from the vibrational displacement is much more difficult, even though some significant progress has been achieved~\cite{SanchezDG}; while it has been a standard setup in the experiment to analyze the mechanical oscillation from the energy spectrum. To this end, we should also study the spring-like mechanical oscillation in KSiNWs from their energy spectrum.

\begin{figure}[htpb]
  \begin{center}
    \scalebox{1.0}[1.0]{\includegraphics[width=\textwidth]{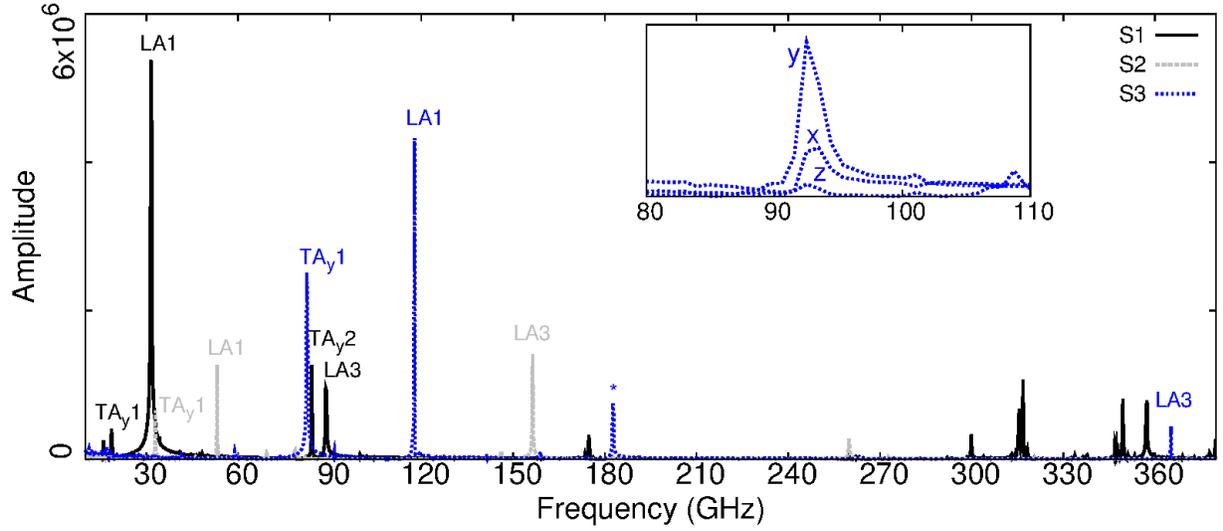}}
  \end{center}
  \caption{The energy spectrum, i.e the Fourier transform of the kinetic energy time history. The first several low-frequency oscillation modes are denoted for all of the three studied samples. With decreasing aspect ratio from S1, S2 to S3, the second transverse mode in $y$ (TA$_{y}$2) undergoes a transition from one-dimensional nanowire behavior into three-dimensional bulk behavior. Inset: Fourier transform of $u_{x}$, $u_{y}$, and $u_{z}$ for S3.}
  \label{fig_fft}
\end{figure}

Fig.~\ref{fig_fft} shows the energy spectrum for all samples, which are actually the Fourier transform of the kinetic energy time history~\cite{JiangJW2012nanotechnology}. Each peak in the figure corresponds to a special oscillation mode in the system. We note that the oscillation mode here is essentially equivalent to the concept of phonon vibration mode in the lattice dynamics theory. There are three low-frequency branches for vibration modes in KSiNWs: one for longitudinal acoustic (LA) mode and two for transverse acoustic (TA) modes. In KSiNW S1, a strong peak locating at $f=31.26$~{GHz} is assigned to the first lowest-frequency LA (LA1) mode. On the left side of this LA1 peak is a smaller peak at $f=18.60$~{GHz} corresponding to the first lowest-frequency transverse mode in $y$ direction (TA$_{y}$1). Both LA1 and TA$_{y}$1 peaks are successfully assigned by comparing the position of these peaks to the frequency extracted from Fig.~\ref{fig_u}. We have also denoted the two higher-frequency modes in $x$ and $y$ directions, i.e LA3 and TA$_{y}$2, at $f=88.69$ and 84.16~{GHz}. The frequency of LA3 is threefold of the frequency for LA1, because the frequency of the LA mode is linearly proportional to the wavelength. Hence the frequency is linearly proportional to its mode index. There are only LA modes with odd index in such quasi-one-dimensional structure with one end fixed and the other end free~\cite{LandauLD}. The frequency of TA$_{y}$2 mode is about fourfold of that for the TA$_{y}$1 mode. It is because of the flexural nature of the transverse modes in one-dimensional rod-like structure of high aspect ratio~\cite{KrishnanA,JiangJW2006,JiangJW2008}, where the frequency is proportional to the square of the mode index.

For KSiNW S2, we also successfully assign the three low-frequency modes, LA1, LA3 and TA$_{y}$1, to the three strongest peaks at $f=53.17$, 156.40, and 32.90~{GHz}, respectively. However, we do not observe a peak corresponding to the TA$_{y}$2 mode, which should occur around $f=131.60$~{GHz}. Similar as S2, we can assign the three low-frequency modes, LA1, LA3 and TA$_{y}$1, to the three strongest peaks in KSiNW S3 at $f=117.78$, 359.54, and 82.49~{GHz}. Similarly, we do not see a peak corresponding to TA$_{y}$2 mode, which should locate around $f=329.96$~{GHz} due to the flexural nature of the transverse mode. However, there is an obvious peak at $f=182.63$~{GHz} (denoted by a star). Inset of the figure shows the Fourier transform for the vibrational displacements $u_{x}$, $u_{y}$, and $u_{z}$. There is a strong peak at $f=92.51$~{GHz} for $u_{y}$, indicating that the resonant peak at $f=182.63$~{GHz} in the energy spectrum should be assigned to a particular oscillation mode in the $y$ direction. It means that this peak corresponds to the TA$_{y}$2 mode in S3, which is contrary to the predicted position $f=329.96$~{GHz} from its one-dimensional flexural nature. This contrary actually reflects the three-dimensional bulk property of the KSiNW S3. As well known, in bulk system, the transverse mode is not flexural anymore. Instead, its frequency is also linear proportional to the wavelength, so its frequency is also proportional to the mode index number. According to this bulk prediction, the frequency of the TA$_{y}$2 mode should be twice of the TA$_{y}$1 mode, i.e $f\approx 164.98$~{GHz}. This value is very close to the peak at $f=182.63$~{GHz} in the energy spectrum. This finding discovers the dimensional crossover (from one-dimensional to three-dimensional) for the transverse mode in KSiNWs with decreasing aspect ratio.

To detect possible effect from the actuation strain $\epsilon_{m}$, we have also performed simulations using $\epsilon_{m}=0.02$. The obtained frequencies of the spring-like mechanical oscillation are compared with that from $\epsilon_{m}=0.01$ in Tab.~\ref{tab_frequency}. There is only slightly enhancement of the frequency by using larger actuation strain. This enhancement effect is weak because of the free boundary condition on the right end, which leads to the absence of the effective strain. It is better to fix both ends to induce effective strain for applications where strong frequency enhancement is desirable.\cite{JiangJW2012nanotechnology}

To summarize, massive MD simulations have been performed to investigate the spring-like mechanical oscillation of KSiNWs with three different aspect ratio. The spring-like oscillation is actuated by the method borrowed from the actuation of a classical mass spring oscillator. The energy spectrum is utilized to analyze the mechanical oscillation of KSiNWs, where the spring-like behavior is detected, accompanying with some other low-frequency oscillation modes. A dimensional crossover is observed for TA$_{y}$ mode in KSiNW samples with aspect ratio decreasing from 3.6 to 2.4.

\textbf{Acknowledgements} The work is supported by the Grant Research Foundation (DFG) Germany.


%

\begin{table*}[t]
\caption{Frequency (GHz) of the spring-like mechanical oscillation in three samples actuated with two different actuation strains $\epsilon_{m}=$ 0.01 and 0.02.}
\label{tab_frequency}
\begin{ruledtabular}
\begin{tabular}{clll}
Sample & S1 & S2 & S3 \\
\hline
$\epsilon_{m}=0.01$ & 15.63 & 26.50 & 58.57 \\
$\epsilon_{m}=0.02$ & 16.21 & 27.66 & 60.08 \\
\end{tabular}
\end{ruledtabular}
\end{table*}

\end{document}